\documentclass[usenatbib,useAMS]{mn2e}

\usepackage{epsf}
\usepackage[dvips]{epsfig}
\usepackage{times}

\begin{document}

\label{firstpage}

\title[Modelling Boo]{Modelling the dynamical evolution of  the Bootes
  dwarf spheroidal galaxy} 

\author[Fellhauer et al.]
{M. Fellhauer$^{1}$ \thanks{Email:
    madf@ast.cam.ac.uk,nwe@ast.cam.ac.uk,vasily@ast.cam.ac.uk},
  M.I. Wilkinson$^{2}$,
  N.W. Evans$^{1}$, 
  V. Belokurov$^{1}$, 
  M.J. Irwin$^{1}$,  \newauthor 
  G. Gilmore$^{1}$, 
  D.B. Zucker$^{1}$,
  J.T. Kleyna$^{3}$ \\
  $^{1}$ Institute of Astronomy, University of Cambridge, Madingley
  Road, Cambridge CB3 0HA, UK \\
  $^2$ Dept.\ of Physics and Astronomy, University of Leicester,
  University Road, Leicester LE1 7RH, UK\\
  $^3$ Institute for Astronomy, 2680 Woodlawn Drive, Honolulu, HI,
  96822, USA}

\pagerange{\pageref{firstpage}--\pageref{lastpage}} \pubyear{2007}

\maketitle

\begin{abstract}
  We investigate a wide range of possible evolutionary histories for
  the recently discovered Bootes dwarf spheroidal galaxy, a Milky Way
  satellite.  By means of $N$-body simulations we follow the evolution
  of possible progenitor galaxies of Bootes for a variety of orbits in
  the gravitational potential of the Milky Way.  The progenitors
  considered cover the range from dark-matter-free star clusters to
  massive, dark-matter dominated outcomes of cosmological simulations.
  For each type of progenitor and orbit we compare the observable
  properties of the remnant after 10~Gyr with those of Bootes observed
  today.  Our study suggests that the progenitor of Bootes must have
  been, and remains now, dark matter dominated.  In general our models
  are unable to reproduce the observed high velocity dispersion in
  Bootes without dark matter.  Our models do not support
  time-dependent tidal effects as a mechanism able to inflate
  significantly the internal velocity dispersion.  As none of our
  initially spherical models is able to reproduce the elongation of
  Bootes, our results suggest that the progenitor of Bootes may have
  had some intrinsic flattening.  Although the focus of the present
  paper is the Bootes dwarf spheroidal, these models may be of general
  relevance to understanding the structure, stability and dark matter
  content of all dwarf spheroidal galaxies.
\end{abstract}


\section{Introduction}
\label{sec:intro}

The dwarf satellite galaxies of the Milky Way have attracted
considerable interest in recent years due to their high apparent mass
to light ratios, which suggest the presence of considerable quantities
of dark matter~\citep[DM:][]{aaronson83,mateo98,kleyna01}.  Given the
apparent absence of dynamically significant DM in globular star
clusters, the dwarf spheroidal galaxies (dSphs) are particularly
valuable for DM studies as they are the smallest stellar systems known
to contain DM~\citep{gilmore07}.  $N$-body simulations of galaxy
formation in a $\Lambda$CDM cosmology predict that a galaxy like the
Milky Way should be surrounded by several hundred low-mass satellite
halos~\citep[e.g.][]{moore99}.  The small number of known dSphs in
the vicinity of the Milky Way has often been cited as a problem for
$\Lambda$CDM and many attempts have been presented in the literature
to explain why the numbers of satellites which formed stars might be
much lower than the total number of substructures.

The past two years have witnessed the discovery of a plethora of new
low-luminosity Milky Way
satellites~\citep{willman05,be06,zu06,be07,walsh07}.  These objects  
probe a previously unexplored regime for galaxies, extending the faint 
end of the galaxy luminosity function by almost four magnitudes.  As
\citet{gilmore07} discuss, these new objects also extend to fainter
magnitudes the apparent bi-modality in the size distribution of
low-luminosity stellar systems, with dSphs exhibiting core radii which
are always more than a factor of four larger than the half-light radii
of the most extended star clusters.  Given that all the dSphs brighter
than $M_{\rm V}=-8$ show evidence of DM, while star clusters appear to
be purely stellar systems, understanding the physical origin of this
size difference may have implications for our knowledge of DM.  It is
therefore of particular importance to determine whether or not the
newly discovered satellites display evidence of DM.  The goal of this
paper is to investigate whether the current observational data for one
of these newly identified satellites can be used to constrain its DM
content. 

\citet{be06} reported the discovery of a faint dwarf galaxy in the
constellation of Bootes.  The object was discovered during a
systematic search for over-densities of stars in the magnitude range
$16 \leq r \leq 22$ in the Sloan Digital Sky
Survey~\citep[SDSS][]{yo00}.  The morphology of the stellar
iso-density contours based on the SDSS data suggested that the system
was quite irregular with some hints of the presence of internal
substructure, possibly indicating that the object was in the process
of tidal disruption. The aim of this paper is to test this hypothesis
and constrain the properties of potential progenitors of this system.

The investigation of the evolution of the dSph galaxies around the
Milky Way by means of $N$-body simulations has a long history.  Many
authors \citep[e.g.][]{jo99,jo02,ma02,read06} try to model the initial
cosmological halos with or without the luminous component inside,
follow their evolution and compare the final results with the
population of dSph galaxies of the Milky Way.  Mostly these studies
focused on the more luminous dwarfs, which were the only known
satellites prior to the recent discoveries.  In this paper we follow a
different approach.  We use $N$-body simulations to investigate the
evolution of possible progenitors of Bootes (henceforth Boo).  Our
goal is to understand the present-day properties of Boo and to
determine whether these can be used to constrain the DM content of
this satellite and of its progenitor.  In \S~\ref{sec:obsprop} we
describe the observed properties of Boo.  \S~\ref{sec:setup}
summarises the initial conditions for our models.
\S~\ref{sec:results} presents the results of our simulations for a
range of possible progenitors and Galactocentric orbits.  Finally, in
\S~\ref{sec:conclus} we assess the relative merits of the various
models and describe the follow-up observations which are required to
distinguish between them.

\section{Observed properties of Boo}
\label{sec:obsprop}

The Bootes dSph is currently located at~\citep{be06,si06}
\begin{eqnarray}
  \label{eq:pos}
  {\rm RA} & = & 14^{\rm h} \, 00^{\rm m} \, 06^{\rm s} \\
  {\rm Dec} & = & +14^{\circ} \, 30\arcmin \, 00\arcsec \\
  D_{\odot}  & = & 62 \pm 3 \ {\rm kpc}
\end{eqnarray}
After correcting for unresolved and faint stars, \citet{be06}
calculated that the total luminosity of Boo is $M_{V,{\rm tot}} =
-5.8$ mag.  Taking a conservative stellar mass-to-light ratio of $M/L
= 2$ this translates into a total luminous mass of
$3.7 \times 10^{4}$~M$_{\odot}$.

\citet{mu06} estimated the bulk radial velocity and line of sight
velocity dispersion of Boo to be 
\begin{eqnarray}
  v_{\rm rad,\odot} & = & +95.6 \pm 3.4 \ {\rm km\,s^{-1}}\\
  \sigma_{\rm los} & = & 6.6 \pm 2.3 \ {\rm km\,s^{-1}}
\end{eqnarray}
based on a sample of seven stars.  More recently, \citet{ma07}
obtained a sample of 30 Boo members and estimated these parameters to
be 
\begin{eqnarray}
  v_{\rm rad,\odot} & = & +99.9 \pm 2.1 \ {\rm km\,s^{-1}}\\
  \sigma_{\rm los} & = & 6.5^{+2.0}_{-1.4} \ {\rm km\,s^{-1}}
\end{eqnarray}
in good agreement with the earlier value.  At present no proper motion
determination for Boo is available.


\citet{be06} compared the colour-magnitude diagram of Boo with that of
the old, metal-poor ([Fe/H]$\sim-2.3$) globular cluster M92 and
concluded that Boo is dominated by a stellar population similar to
that of M92, although slightly younger and more metal poor.  No
evidence of young or intermediate age populations in Boo has yet been
found and no traces of HI gas have been detected~\citep{ba06}.  Based
on the equivalent widths of the Mg lines in the spectra of their seven
Boo members, \citet{mu06} estimated the metallicity of the system to be
$[\rm Fe/H] \approx -2.5$, which would make Boo the most metal-poor
Milky Way dwarf spheroidal discovered to date. \citet{ma07} found a
somewhat higher metallicity of $[\rm Fe/H] \approx -2.1$ based on the
equivalent widths of the CaII near-infrared triplet (CaT) lines from a
larger sample of 19 stars.  The origin of the discrepancy between these
two measurements is unclear, but could be related to the difference in
the calibration of the empirical estimators for [Fe/H] from the Mg and
Ca line-widths.  Prior to the work of Batagglia et al.\ 2007, neither
of these measures was calibrated below [Fe/H]$\sim -2$ (Rutledge et al.\
1997; see also the discussion in Koch et al.\ 2007).  Battaglia et al.\
2007 compared abundances derived from high resolution spectra with
abundances for the same stars derived from low resolution spectra of
the CaT and showed that, although the CaT calibration holds until at least
$[\rm Fe/H] = -2.5$, offsets of derived [Fe/H] of $\approx \pm$0.2 dex can
occur depending on how the measurement and calibration of the low
resolution spectra is done.  However, despite the difference in these
estimates the robust implication is that the stars in Boo are both old
and metal-poor.

In Fig.~\ref{fig:cont}, we present new stellar iso-density contours
based on deeper photometric data obtained with SuprimeCam mounted on
the Subaru telescope.  In contrast to the SDSS contours, the Subaru
data suggest that Boo has a regular internal morphology, although some
internal substructure may still be present (see Fig.~\ref{fig:cont}).
We note however, that some of the distortion seen in the inner
contours is due to the presence of a very bright star in the field of
view.  It is thus possible that the internal morphology of Boo is
quite regular.  Both sets of contours exhibit elongation along an axis
of roughly constant RA with an hint of an S-shape in the contours - we
will use this later when choosing possible orbital paths for Boo.

\section{Setup}
\label{sec:setup}

\begin{figure}
  \centering 
  \epsfxsize=7cm \epsfysize=7cm \epsffile{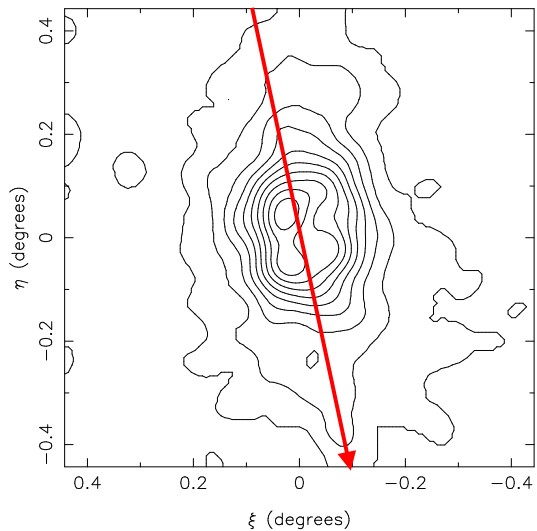}
  \caption{Contours of the Bootes dwarf satellite constructed from a
  stellar isopleth map of objects selected to occupy the main locus of
  Boo members in the CMD.  Contours start at $2\sigma$ above the
  background in steps of $2\sigma$.  The photometric data were
  obtained using SuprimeCam mounted on the Subaru telescope.  The
  straight (red) line shows a possible projected orbital path of
  Boo.}  \label{fig:cont} 
\end{figure}

Based on the observed data discussed in \S~\ref{sec:intro}, there are
four plausible assumptions which we can make in order to find possible
orbits and progenitors of Boo:
\begin{enumerate}
\item The iso-density contours of Boo are elongated approximately in
  the direction of constant RA and appear to show the (S-shaped)
  on-set of tidal tails.  The tail-like features constrain the size of
  the luminous object and may be used to determine the projected
  orbital path.  
\item The measured line-of-sight velocity dispersion is of
  order $7$~km\,s$^{-1}$, which constrains the total mass of the
  object at the present epoch.
\item The scale-length (half-light radius) of the luminous matter is
  approximately $13^{\prime}$ (corresponding to $230$~pc at the
  distance of Boo) and the central surface brightness is of the order
  of $\mu_{0} = 28$~mag\,arcsec$^{-2}$.
\item The internal substructure seen in the contours of
  Fig.~\ref{fig:cont} is not real, but instead is likely to be an
  artifact of the sparse photometric data.
\end{enumerate}

In choosing the properties of our Boo progenitors below, we will make
use of some or all of these assumptions.  We emphasise that these are
assumptions for the present analysis, and their validity is amenable
to future observational test.

\subsection{Galactic Potential}
\label{sec:potential}

In our study, we model the Galactic potential analytically using a
logarithmic halo of the form
\begin{eqnarray}
  \label{eq:halo}
  \Phi_{\rm halo} & = & \frac{1}{2} v_{0}^{2} \ln \left( r^{2} +
    d^{2} \right),
\end{eqnarray}
with $v_{0} = 186$~km\,s$^{-1}$ and $d = 12$~kpc. Our choice of a
spherical halo rather than a prolate or oblate model is motivated by
our recent study of the tidal tails of the Sagittarius dSph, which
showed that in the region probed by the debris from Sagittarius, the
gravitational potential has to be close to spherical \citep{fe06}.  We
model the disc as a Miyamoto-Nagai potential \citep{mn75} of the form 
\begin{eqnarray}
  \label{eq:disc}
  \Phi_{\rm disc}(R,z) & = & - \frac{G M_{\rm d}} { \sqrt{R^{2} + \left(
        b + \sqrt{z^{2}+c^{2}} \right)^{2}}},
\end{eqnarray}
with $M_{\rm d} = 10^{11}$~M$_{\odot}$, $b = 6.5$~kpc and $c =
0.26$~kpc (where $R$ and $z$ are cylindrical coordinates).  The bulge
is represented by a \citet{he90} potential
\begin{eqnarray}
  \label{eq:bulge}
  \Phi_{\rm bulge}(r) & = & - \frac{G M_{\rm b}} {r+a},
\end{eqnarray}
using $M_{\rm b} = 3.4 \times 10^{10}$~M$_{\odot}$ and $a = 0.7$~kpc. 

\subsection{Possible Orbits}
\label{sec:orbit}

As a starting point, we take the radial velocity of Boo at its present
position to be $v_{\rm rad} = 100$~km\,s$^{-1}$ and its distance from
the sun to be $D_{\odot} = 60$~kpc.  We use assumption (i) above to
constrain the projected orbital path near the satellite.  Furthermore,
the shape of the stellar distribution near the on-set of the putative
tidal tails suggests that the direction of motion along this path is
likely to be from the top of Fig.~\ref{fig:cont} to the bottom.  It is 
well-established that the tidal tails of a disturbed stellar system
lie close to the orbital path of the system and that the leading arm
should be closer to the Galactic Centre than the object while the
trailing arm should be more distant~\citep[see e.g.][]{combes99}.  In
Fig.~\ref{fig:cont}, the Galactic Centre lies approximately in the
direction of the lower left corner of the figure which motivates our
assumption regarding the direction of motion.

With the present-epoch radial velocity and projected path fixed, we
use a simple point mass integrator to determine the forward and
backward (in time) orbits for given pairs of assumed proper motions.
Orbits which move along the assumed projected path are regarded as
possible orbits of Boo.  In Table~\ref{tab:orbit} we give a selection
of such orbits.  They span a wide range of perigalactica, apogalactica
and eccentricities and include orbits which have very close approaches
to the Galactic Centre as well as orbits which spend most of their
time in the outer Milky Way halo.  Thus, despite restricting our
models to the orbital path implied by assumption (i) we are still able
to access the whole parameter space of possible eccentricities, peri-
and apogalactic distances, even if assumption (i) turns out to be
wrong.

We choose orbit (4) from Table~\ref{tab:orbit} as the standard one for
our simulations.  This choice is more or less arbitrary and is
justified only by the fact that it not in any sense an extremal
orbit: it neither gets very close to the Galactic Centre nor is it an
orbit with a large apogalacticon.  It also has a moderate
eccentricity. 

We also run models for orbit (3), which has an eccentricity more
similar to the eccentricities of sub-halos seen in cosmological
simulations \citep[e.g.][]{gi98} and one extreme model choosing orbit
(2).  

In our study, the simulation time is kept fixed at $10$~Gyr.
This choice is as arbitrary as the choice of orbit. It is likely that
there is a trade-off between the choice of simulation time and the
choice of orbit. For example, placing an object on an orbit with a
close perigalacticon but only a short simulation time could lead to a
similar remnant system as that produced by a progenitor farther out in
the halo but whose evolution was followed for a longer time. However,
while this would add another dimension to the (already vast)
parameter-space of this problem, it would not affect our general
conclusions about the properties of plausible progenitors.

\begin{table}
  \centering 
  \caption{The sample of the possible orbits of Boo simulated in this
    study.  The first two columns give the proper motions of the orbit
    (the radial velocity at the present location of Boo is assumed to
    be $100$~km\,s$^{-1}$ in all cases).  The third and fourth column
    give the peri- and apogalacticon distances and the last column
    gives the eccentricity of the orbit.}  
  \label{tab:orbit} 
  \begin{tabular}{cccccc} \hline
    name & $\mu_{\alpha}$ & $\mu_{\delta}$ & $R_{\rm peri}$ & $R_{\rm
      apo}$ & $e$ \\ 
    & [mas\,yr$^{-1}$] & [mas\,yr$^{-1}$] & [kpc] & [kpc] & \\ \hline 
    (1) & -0.53 & -0.62 &  1.8 & 66.2 & 0.95 \\ 
    (2) & -0.54 & -0.70 &  4.7 & 66.2 & 0.87 \\ 
    (3) & -0.58 & -0.90 & 14.8 & 67.2 & 0.64 \\ 
    (4) & -0.63 & -1.20 & 36.9 & 76.7 & 0.35 \\ 
    (5) & -0.66 & -1.40 & 48.8 & 104.3 & 0.36 \\ \hline 
  \end{tabular}
\end{table}

\subsection{Possible Progenitors}
\label{sec:inimod}

In order to determine the range of plausible progenitors for the
satellite, we must now have a closer look at the properties of the Boo
dwarf at the present epoch.  If our assumption (i) regarding an
interpretation of the morphology of the outer iso-density contours as
the on-set of tidal tails is correct, this would imply an approximate
tidal radius for Boo of about $250$~pc.  In our analytical model of
the Milky Way, the enclosed mass ($M_{\rm MW}$) at a distance $D_{\rm
  GC} \approx 60$~kpc is roughly $6 \times 10^{11}$~M$_{\odot}$.
Using the Jacobi limit for a satellite on a circular orbit
\citep[][their eq.~7-84]{bi87}: 
\begin{eqnarray}
  \label{eq:rtidal}
  r_{\rm tidal} & = & \left( \frac{M_{\rm sat}} {3 M_{\rm MW}}
  \right)^{\frac{1}{3}} D_{\rm GC}
\end{eqnarray}
we find that the mass within the tidal radius of Boo is approximately
$7 \times 10^{4}$~M$_{\odot}$, comparable to the observed luminous
mass in Boo.  While one should not over-interpret the closeness of the
agreement given the crudeness of the above estimates, this suggests
that it is possible to find progenitors for Boo which do not require
that Boo be dark matter dominated.  However, such models must also be
consistent with the observed internal velocity dispersion, which
provides tighter constraints.  To illustrate this we now adopt a
\citet{pl11} model for the satellite:
\begin{eqnarray}
  \label{eq:plummer-rho}
  \rho_{\rm pl} (r) & = & \frac{3 M_{\rm pl}} {4 \pi R_{\rm pl}^{3}}
  \left( 1 + \frac{r^{2}}{R_{\rm pl}} \right)^{-5/2}
\end{eqnarray}
with a Plummer radius ($R_{\rm pl}$; which corresponds to the
half-light radius) of $200$~pc. The formula
\begin{eqnarray}
  \label{eq:plummer-los}
  \sigma_{\rm los}(0) & \approx & 2.52 \times \sqrt{ \frac{M_{\rm
        sat} [\rm 10^{7}M_{\odot}]} {R_{\rm pl} [\rm kpc]}} \ [\rm
  km\,s^{-1}] 
\end{eqnarray}
yields a central line-of-sight velocity dispersion for such an object
of only $0.5$~km\,s$^{-1}$.  This is in clear disagreement with the
measured velocity dispersion of Boo.  

With this background, we now explore the possibilities further.
We first search for a dark matter-free (Plummer model) progenitor of
Boo as our case A, to assess how well it can reproduce the observed
morphology, keeping in mind our expectation that it will not be able
to reproduce the observed velocity dispersion.  In this model, the
progenitor of Boo is similar to a 'star cluster' in the sense that it
contains no dark matter, but it is unlike any known star cluster: its
half-light radius is significantly larger than any observed star
cluster.  It resembles a dwarf galaxy like a tidal dwarf galaxy
without dark matter content.

Alternatively, we can use Eq.~\ref{eq:plummer-los} as a mass estimator
and compute the mass of a model with a line-of-sight velocity
dispersion of $7$~km\,s$^{-1}$ and a scale-length of $200$~pc.  This
calculation yields a total mass of $1.5 \times 10^{7}$~M$_{\odot}$.
Inserting this value into Eq.~\ref{eq:rtidal} we find that the tidal
radius of Boo at its current location would be $1.2$~kpc (or $1^{o}$).
Although this is much larger than the apparent size of Boo, given the
low surface brightness of the system it is possible that the stellar
distribution extends significantly further than the contours of
Fig.~\ref{fig:cont} might suggest.  We therefore search for a possible
progenitor using a single-component Plummer model as our case B, or
mass-follows-light model.  We note that if this model is correct, it
would mean that the elongation of the stellar iso-density contours is
intrinsic to Boo rather than being the on-set of tidal tails.  We do
not attempt to reproduce this elongation in our simulations, as this
would add two more dimensions (angles of the initial orientation) to
the space of free parameters in which we are searching for a
progenitor, without changing our conclusions significantly.  Although
in this model the elongation is not tidally induced, we nevertheless
use orbit (4) for this simulation to ease comparison with our
'star-cluster' model.  As we discuss below, the results are
qualitatively similar for a progenitor system placed on more extreme
orbits.  In a more general sense our model B could be regarded as a
cored halo model with the luminous component having the same
scale-length as the halo.    

More fundamentally, there is no a priori reason why the stellar
distribution should extend to the physical tidal radius.  A similar
argument applies to our models C1 and C2, in which we adopt a more
elaborate, two-component representation of a dark matter-dominated
satellite.  We use a \citet{he90} profile: 
\begin{eqnarray}
  \label{eq:hernquist}
  \rho_{\rm H} (r) & = & \frac{M_{\rm H}} {2 \pi} \frac{r_{\rm
      sc}} {r} \frac{1} {(r + r_{\rm sc})^{3}}
\end{eqnarray}
to represent the luminous matter (where $M_{\rm H}$ is the total
luminous mass and $r_{\rm sc}$ is the scale-length) embedded in a halo
of the form 
\begin{eqnarray}
  \label{eq:nfw}
  \rho_{\rm NFW} (r) & = & \rho_{0} \frac{r_{\rm s}} {r \left( 1 +
      \frac{r}{r_{\rm s}} \right)^{2}},
\end{eqnarray}
\citep[][henceforth NFW]{nfw96} where $\rho_{0}$ is the characteristic 
density and $r_{\rm s}$ is the scale-length.  Due to the significant
amount of mass at large radii in this model, such a progenitor will
not develop tidal tails within the observed field of Boo.  As in model
B, therefore, this scenario requires the initial stellar distribution
to be elongated.  As before, we do not attempt to reproduce this
elongation.  We investigate two extreme sub-cases of scenario C (the
extended DM scenario): first (C1) a luminous sphere embedded in a DM
halo with the same scale-length and secondly (C2) a model with a very
massive and extended DM halo whose scale-length is larger than the
extent of the luminous matter.  The parameters of both halos in our C
models are broadly consistent with the outcome of large-scale
cosmological $\Lambda$CDM simulations: in particular their scale radii
and concentrations lie within the (broad) range expected for
cosmological halos of these masses \citep{ji00}.

In order to investigate further the possible relevance of putative
tidal tails in dark matter-dominated models, we also consider a
scenario in which the progenitor of Boo is on a quite extreme
Galactocentric orbit.  In this model, the satellite is initially
deeply embedded in a much larger halo.  To ensure that the luminous
matter becomes tidally distorted, this system has to approach the
Galactic Centre sufficiently closely that its tidal radius at
perigalacticon shrinks to about $200$~pc.  To be consistent with the
observed high velocity dispersion of Boo, the total mass of the object
interior to this radius is $1.5\times 10^{7}$~M$_{\odot}$.  Using
Eq.~\ref{eq:rtidal} and computing the enclosed mass of the Galaxy
using the integrated forms of Eqs.~\ref{eq:halo}--\ref{eq:bulge} we
see that in this case the perigalacticon of Boo's orbit must be about
$5$~kpc.  During the perigalactic approach, the outer halo of Boo will
become unbound and even the luminous part will be tidally shocked,
developing tidal tails.  As it moves to larger radii, the tidal radius
expands to its much larger value at the present epoch and the
residual outer halo, which did not have time to escape, becomes
bound again.  We call this our case D (or Dark Matter \& tails model).
For this model we use orbit (2) from Table~\ref{tab:orbit}.

Since our goal is a general analysis of possible histories of the Boo
dSph galaxy, our parameter-space survey also included models in which
the satellite is now in the final stages of disruption and
dissolution.  In order to be so strongly affected by tides, such
models are necessarily of lower initial mass than the models which
retain significant mass until late times.  We found that such models
could not reproduce the observed high velocity dispersion while
simultaneously satisfying the constraint of the scale-length being
$13$~arcminutes (assumption iii).  This conclusion holds using either
single-component Plummer models or two-component, initially dark
matter-dominated models as progenitors.  That is, models which are not
dark matter dominated but are tidally disrupting are not, in general,
able to reproduce the observed velocity dispersion measured in Boo. 
This conclusion is of more
general relevance for our understanding of the dark matter content of
dSph galaxies. 

This adds weight to our assumption that the internal photometric
substructure in Boo is not real.  While internal substructure is a
possible signature of tidal disruption, it is rapidly erased in the
interior of a massive, bound system~\citep[see e.g][]{fe05} except in
certain circumstances~\citep[see e.g.][]{kleyna03}.  The failure of
disrupting models to reproduce the velocity dispersion of Boo suggests
that the observed substructures are either kinematically very cold, or
are due to noise in the photometric data set.  It also means that the
visible elongation of Boo is intrinsic to the progenitor of Boo and
not tidally induced.  Regarding the possible S-shape of the contours
it can only be speculated that an initially flattened system, which is
at least partly rotationally supported, might feel a tidally induced
torque which bends the contours into an S-shape, even though the
progenitor is deeply embedded in a DM halo \citep[see e.g.][]{ma01}.
Given the current sparse nature of the observational data for Boo,
such an investigation is not warranted in the context of the present
paper. 


We also investigated a more speculative scenario in which Boo had a
recent encounter with the Sagittarius dwarf galaxy.  In the absence of 
an observed proper motion, it is possible to find an orbit for Boo
which would produce such a close passage around $340$ Myr ago.  The
proper motion of Boo in this case would be $\mu_{\alpha \cos \delta}
= +0.034$~mas\,yr$^{-1}$, $\mu_{\delta} = -1.024$~mas\,yr$^{-1}$.
However, the remnant of Boo in this case is no more similar to the
observed object than in the other scenarios we have considered and so
the extra assumption of a two-body encounter with Sagittarius is not
warranted.  

For completeness, we also investigated a more exotic ansatz, namely
'star cluster' models (i.e.\ models without dark matter) which have a
very massive black hole at their centre.  As anticipated, even in the
models which contained the most massive black hole which still led to
an undisrupted object (in this case the central black hole was five
times more massive than the stellar remnant) we could not reproduce
the high velocity dispersion, even though the black hole-induced
velocity dispersion dominated the bulk of the remnant.

\subsection{Simulations}
\label{sec:sim}

The modus operandi of our search for possible progenitors is the
following:
\begin{enumerate}
\item We adopt a possible orbit for Boo.  For most of our simulations
  this is our reference orbit (4) from Table~\ref{tab:orbit}.
\item We use a simple point-mass integration to trace this orbit
  backwards for $10$~Gyr.
\item We insert the model for our possible progenitor at this starting
  position.  In each model, each component of the model is represented
  by $10^{6}$ particles.
\item We simulate the model forward in time until the present epoch
  using the particle-mesh code Superbox \citep{fe00}.
\item We analyse the final model and compare it to the observations.
  If there is a mismatch, we alter the initial properties of the
  progenitor (scale-length, mass, etc.) accordingly and start again
  with step (iii).
\end{enumerate}

\begin{figure}
  \centering \epsfxsize=7cm \epsfysize=7cm \epsffile{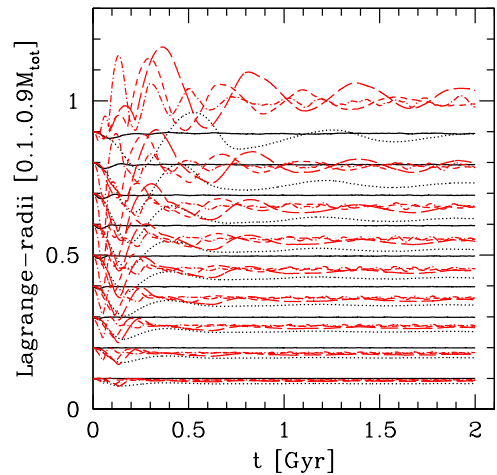}
  \caption{Lagrangian radii of our initial models evolved in isolation
  to demonstrate their stability.  The plot shows the scaled
  Lagrangian radii ($10$\%..$90$\% of the total mass).  The radii are
  scaled by their initial value and multiplied by the mass-fraction
  they represent.  This results in initially equidistant radii
  independent of the choice of the initial profile, but also that
  the relative deviation of the $90$\% radius is displayed nine times
  larger than the deviation of the $10$\% radius.  Dotted line is
  model A, solid line is model B, short dashed line is model C1, long
  dashed line is model C2 and dashed-dotted line represents model D.
  The additional models on orbit (3) are omitted for clarity.  All
  models show an initial adaptation to the grid-based treatment.  The
  additional luminous component for the models C1--D results in a
  slight contraction of the inner parts of the DM halo.  Also an 
  expansion of the outermost radius, due to the cut-off radius is
  visible.  But after this short initial period the radii stay
  constant for the rest of the simulation time.}
  \label{fig:lagrad}
\end{figure}

For all our models, we check for stability by evolving them first in
isolation.  The time evolution in isolation of the Lagrangian radii of
some of our models are shown in Fig.~\ref{fig:lagrad}.  For the
single-component Plummer models, the distribution function is
used to generate the velocities.  For the combined models we use the
Jeans equation~\citep[see e.g.][]{bi87} 
\begin{eqnarray}
  \label{eq:jeans}
  \sigma_{r,{\rm i}}^{2}(r) & = & \frac{1} {\rho_{\rm i}(r)}
  \int_{r}^{rc} \frac{G M_{\rm tot}(r^{\prime})} {r^{\prime 2}}
  \rho_{\rm i}(r^{\prime}) {\rm d}r^{\prime},
\end{eqnarray}
and the Maxwellian approximation~\citep[e.g.][]{he93} to generate the
velocities.  It is known \citep{ka04} that this approximation leads to
a small core in the centre and a velocity anisotropy in the outer
parts, i.e.\ enhancing the mass-loss of the outer shells.  However, we
note the addition of a luminous component to our models influences the
DM distribution in the inner parts, making them more concentrated
\citep{kat04}.  This is clearly visible in Fig.~\ref{fig:lagrad} and
below in Fig.~\ref{fig:time}.  There we show that despite we use the
Maxwellian approximation our DM haloes are cuspy in the centre and not
cored even after $10$~Gyr of evolution in a tidal field.  A cored
profile would indeed have a more impulsive response (due to the longer
internal crossing time) to the tidsal field and henceforth a larger
mas-loss also in the inner parts, but we do not see this effect in our
models.  With regard to the incorrect treatment of the outer parts we
note that we truncate our initial distribution at a radius which is
only slightly larger than the tidal radius of Boo today, thereby
reducing the initial mass-loss of the outer parts.  A halo which
extends to its virial radius ($r_{200}$, a quantity which is an order
of magnitude larger than the tidal radius at first perigalacticon)
would suffer strong mass-loss during its first peri-Galactic passage.
Thus, the region from which we expect enhanced mass-loss due to the
Maxwellian approximation would be removed by tides at early times.  By
starting with a smaller initial truncation radius, we negate the
impact of this mass loss on the evolution of our models. 

\section{Results}
\label{sec:results}

In this section, we describe the best-fitting models of each type
(A-D) introduced in \S\ref{sec:inimod}.  We show their initial
parameters (relevant for the numerical setup) as well as the computed
values of the virial radius, concentration and maximum circular
velocity (relevant for comparison with other studies) in
Table~\ref{tab:models}.  The results of the Boo models at the present
epoch, especially the inferred central mass-to-light ratios of Boo,
are shown in Table~\ref{tab:res}.

\begin{table*}
  \centering
  \caption{Initial conditions for the best fitting models of all six
    cases.  Shown are the basic parameters for each model and each
    component: initial mass, scale-length and cut-off radius.
    Furthermore we denote the virial radius ($r_{\rm vir} = r_{200}$,
    except for the A-cases: $r_{\rm vir} = -GM^{2}/4E$),
    the concentration $c=r_{\rm vir}/r_{\rm s}$, the mass at the
    virial radius, the maximum rotation velocity and finally the orbit
    (numbers refer to orbits in Table\protect\ref{tab:orbit}).} 
  \label{tab:models}
  \begin{tabular}{cccccccccc} \hline
    case & model & mass & $r_{\rm s}$ & $r_{\rm cut}$ & $r_{\rm vir}$
    & $c$ & $M(r_{\rm vir})$ & $v_{\rm c,max}$ & orbit \\
         &       & [$M_{\odot}$] & [pc] & [pc] & [kpc] & &
         [M$_{\odot}$] & [km\,s$^{-1}$] & \\ \hline
    A    & Plummer & $8.0 \times 10^{5}$ &  202 &  500 & 0.34 & 1.7 &
         $5.1 \times 10^{5}$ & 4.1 & (4) \\ \hline
    B    & Plummer & $1.6 \times 10^{7}$ &  200 & 2000 & 1.4 &  7.0 &
         $1.6 \times 10^{7}$ & 18.5 & (4) \\ \hline
    C1   & NFW     & $4.5 \times 10^{7}$ &  300 & 1200 & 12 & 40 &
         $1.5 \times 10^{8}$ & 13.1 & (4) \\
         & Hernq.  & $3.0 \times 10^{4}$ &  300 &  300 &    & \\ \hline
    C2   & NFW     & $3.0 \times 10^{8}$ & 1000 & 2500 & 25 & 25 &
         $1.3 \times 10^{9}$ & 22.7 & (4) \\
         & Hernq.  & $4.0 \times 10^{4}$ &  250 &  500 &    & \\ \hline
    D    & NFW     & $1.25 \times 10^{8}$&  250 & 1000 & 18 & 72 &
         $5.1 \times 10^{8}$ & 23.9 & (2) \\
         & Hernq.  & $5.0 \times 10^{4}$ &  250 &  400 &    & \\ \hline
    A-3  & Plummer & $3.5 \times 10^{6}$ &  178 &  500 & 0.3 & 1.7 &
         $2.2 \times 10^{6}$ & 9.2 & (3) \\ \hline
    B-3  & Plummer & $2.0 \times 10^{7}$ &  200 & 1000 & 2.6 & 2.6 &
         $2.0 \times 10^{7}$ & 20.7 & (3) \\ \hline
    C1-3 & NFW     & $5.0 \times 10^{7}$ &  300 & 1000 & 13 & 45 &
         $2.0 \times 10^{8}$ & 8.0 & (3) \\
         & Hernq.  & $5.0 \times 10^{4}$ &  300 &  500 &    & \\ \hline
    C2-3 & NFW     & $6.0 \times 10^{8}$ & 1000 & 2500 & 32 & 32 &
         $2.8 \times 10^{9}$ & 61.0 & (3) \\
         & Hernq.  & $5.0 \times 10^{4}$ &  250 &  500 &    & \\ \hline
   \end{tabular}
\end{table*}

\begin{figure}
  \centering
  \epsfxsize=4.1cm
  \epsfysize=4.1cm
  \epsffile{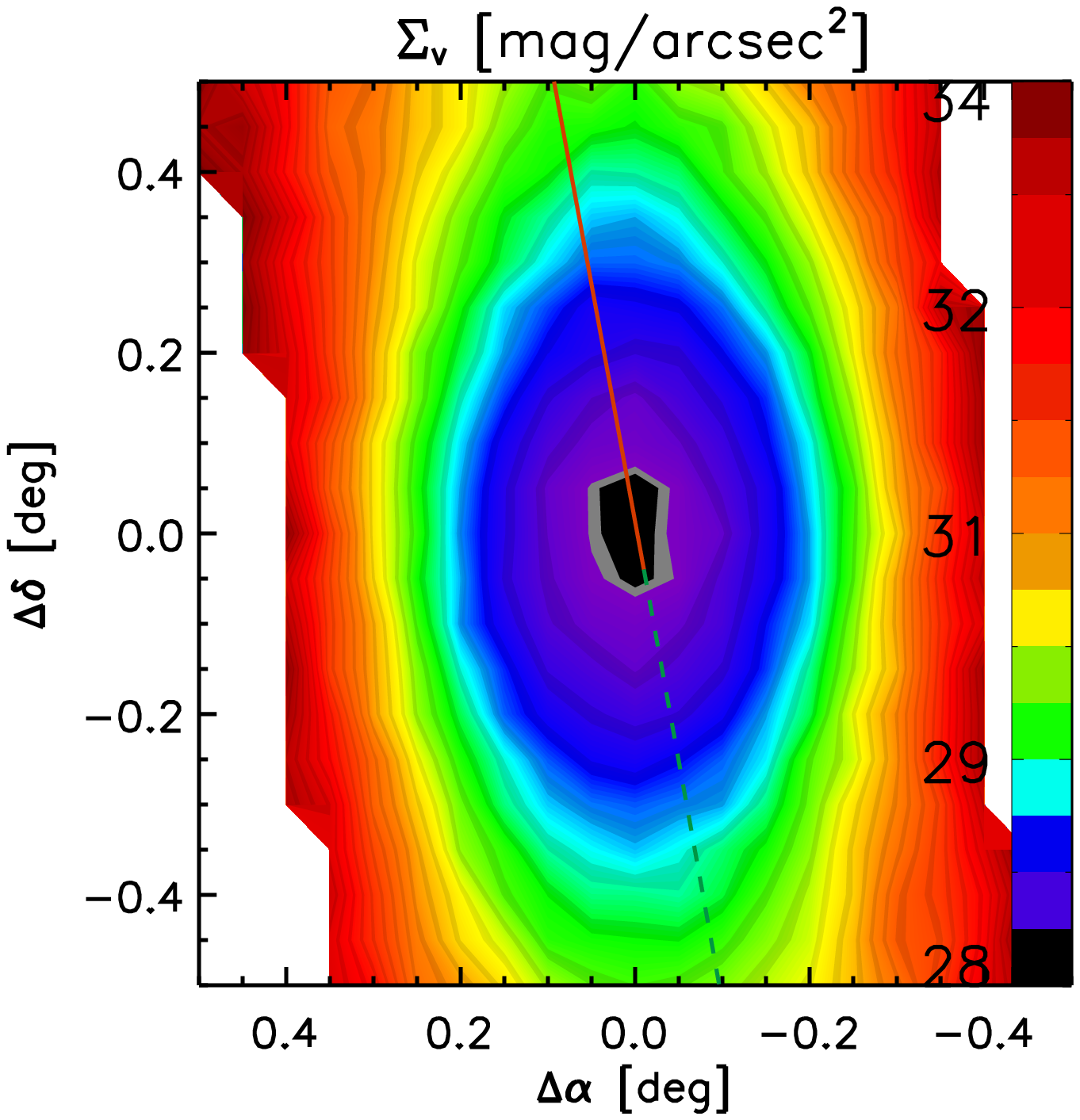}
  \epsfxsize=4.1cm
  \epsfysize=4.1cm
  \epsffile{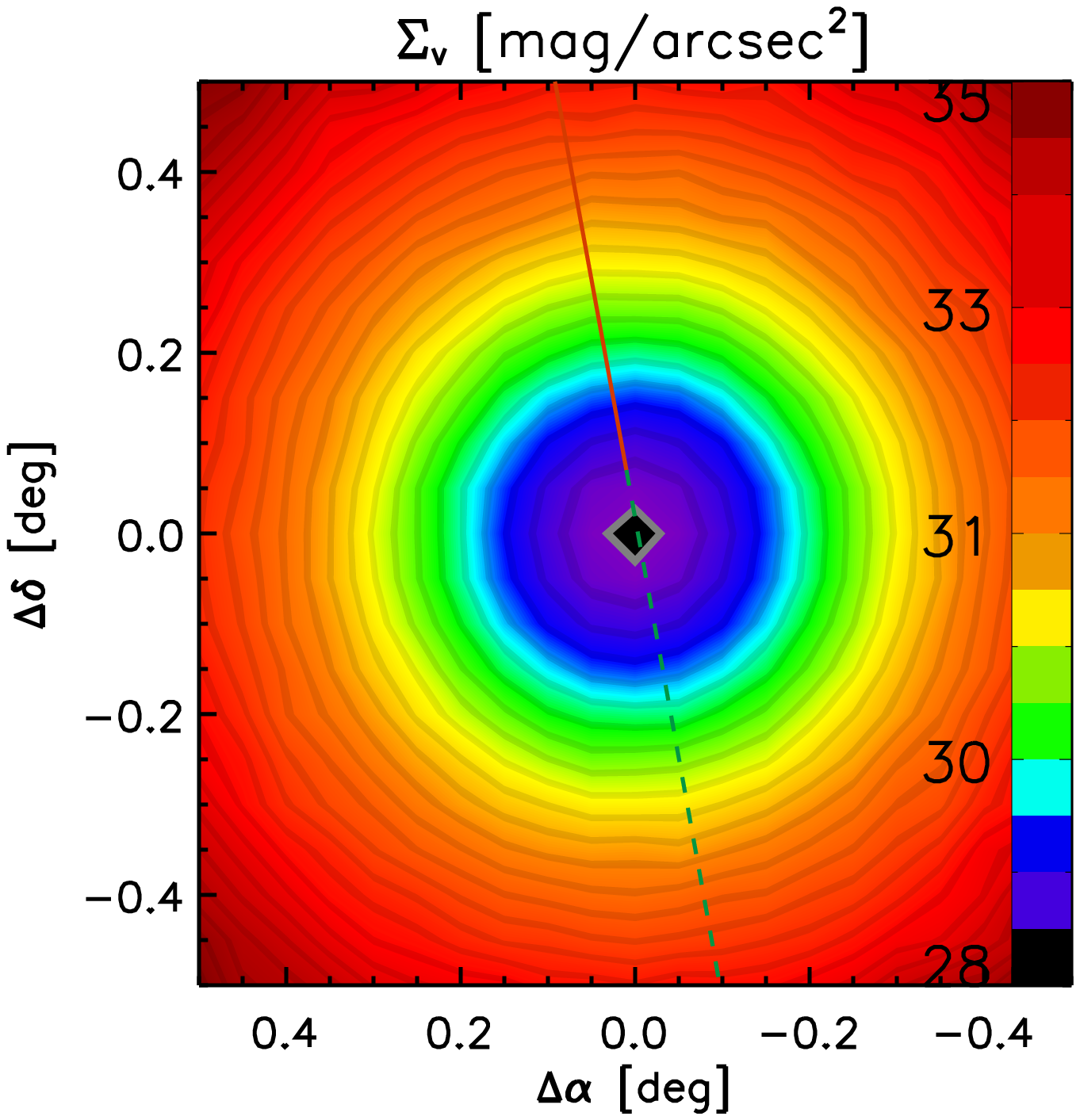}
  \epsfxsize=4.1cm
  \epsfysize=4.1cm
  \epsffile{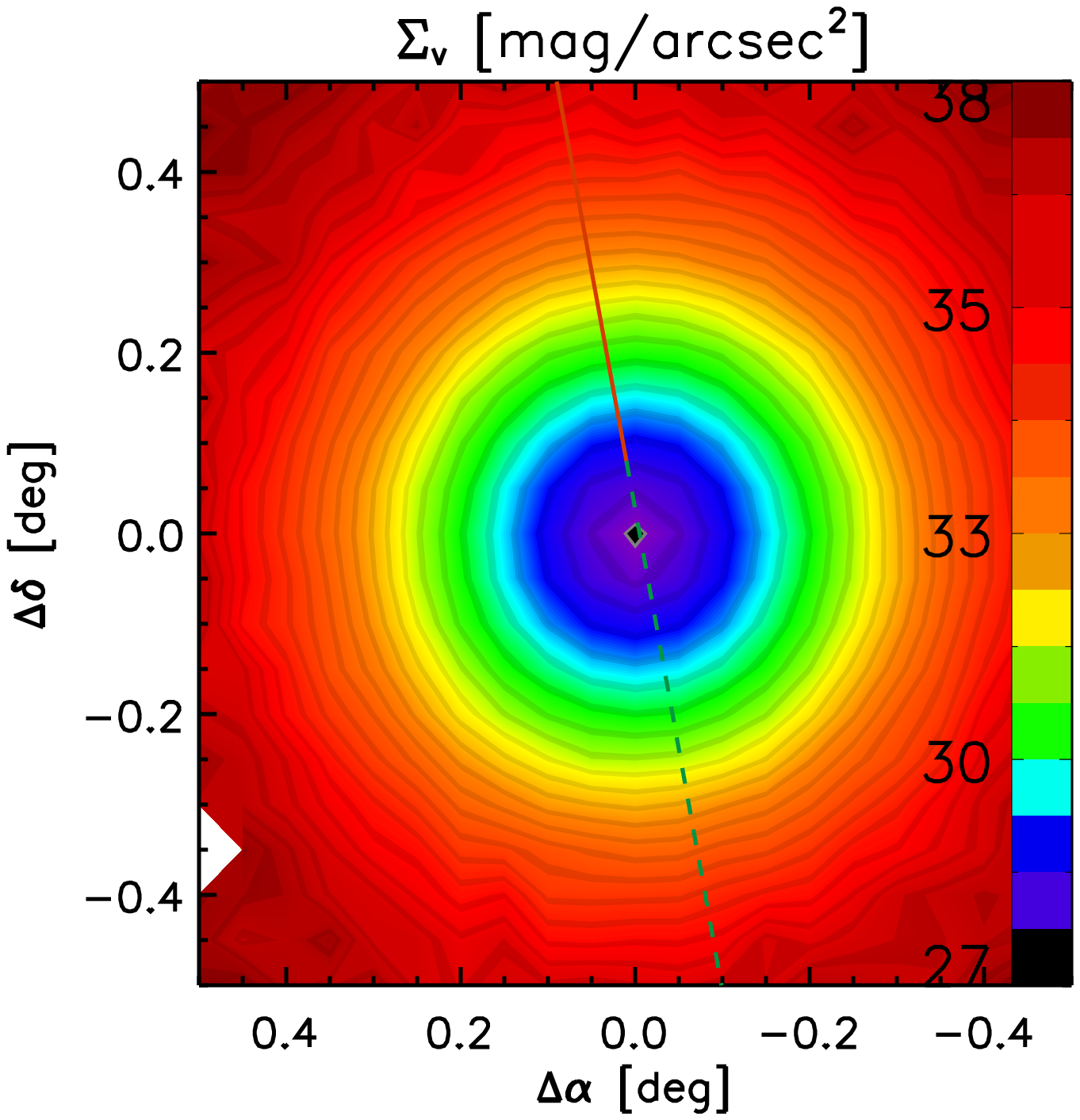}
  \epsfxsize=4.1cm
  \epsfysize=4.1cm
  \epsffile{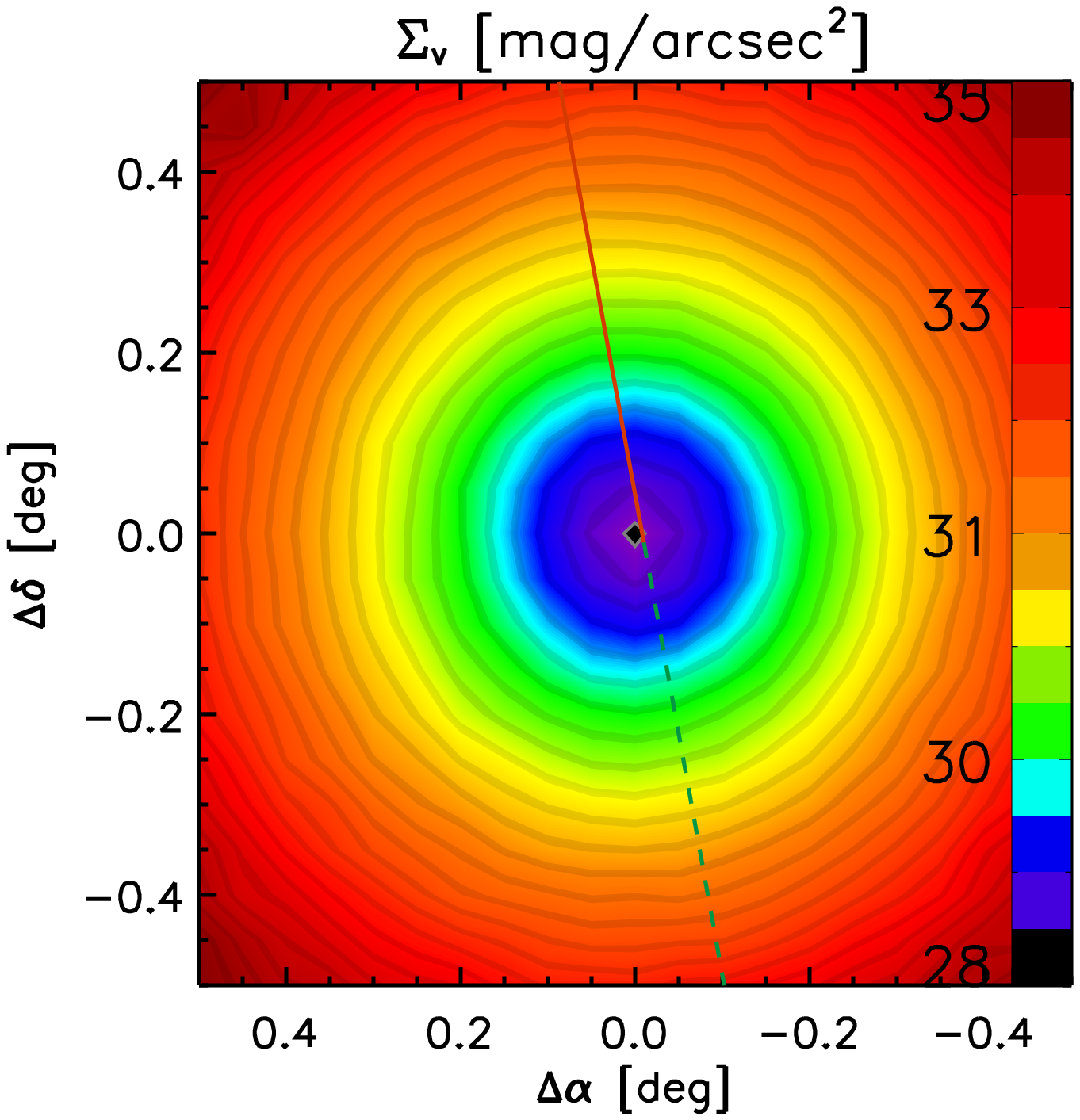}
  \epsfxsize=4.1cm
  \epsfysize=4.1cm
  \epsffile{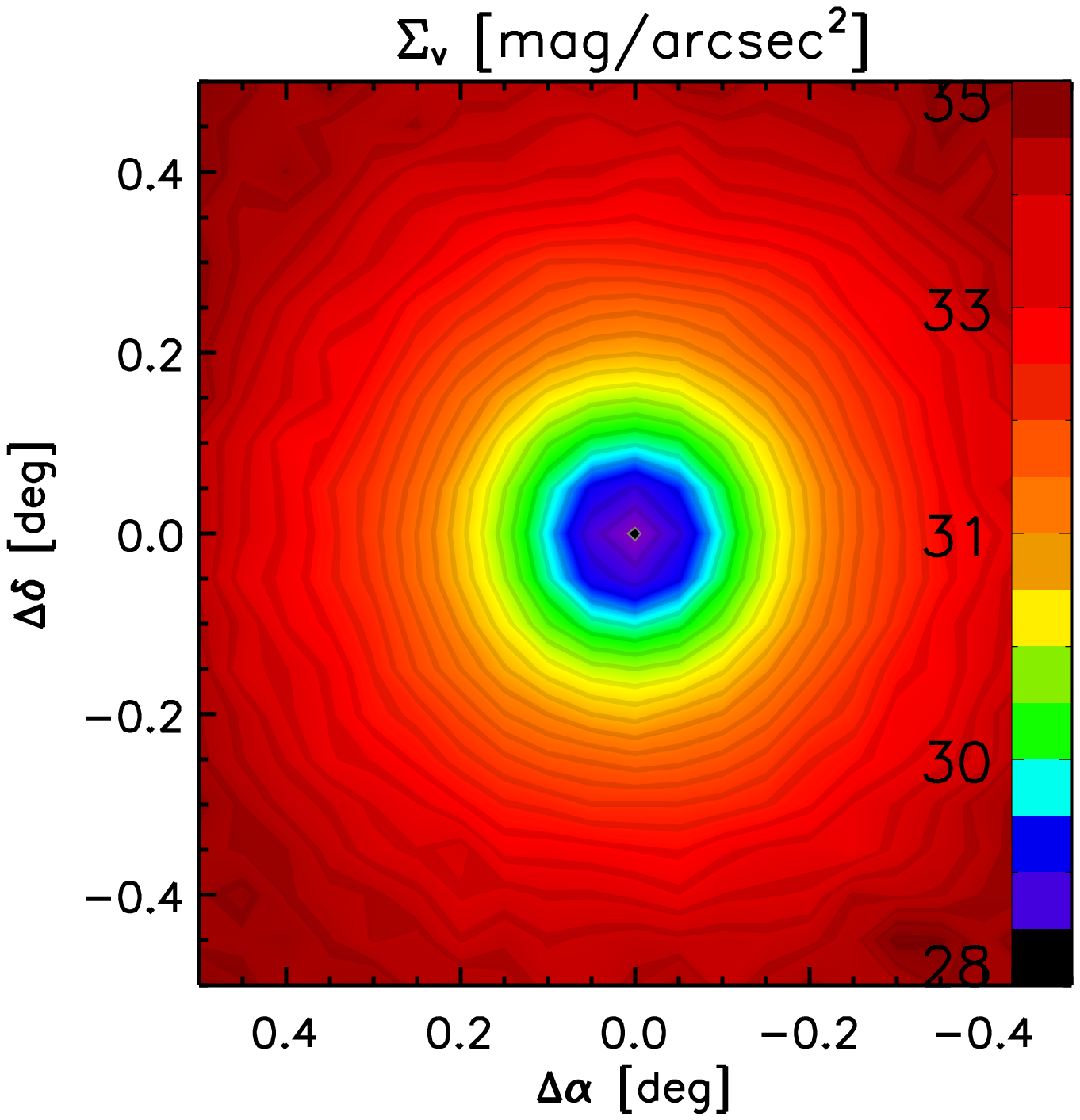}
  \caption{Surface brightness distribution of our models at the
    present epoch.  \textbf{Top left:} case A, star cluster model;
    \textbf{top right:} case B, mass (stellar plus dark) follows light
    model; \textbf{middle left:} case C1, two-component, dark matter
    dominated model with halo of the same scale-length; \textbf{middle
      right:} case C2, 2-component, dark matter dominated with
    extended halo; \textbf{bottom:} case D, 2-component, dark
    matter dominated with tails.  Models on orbit (3) are omitted.} 
  \label{fig:surf-pix}
\end{figure}

\begin{figure}
  \centering
  \epsfxsize=4.1cm
  \epsfysize=4.1cm
  \epsffile{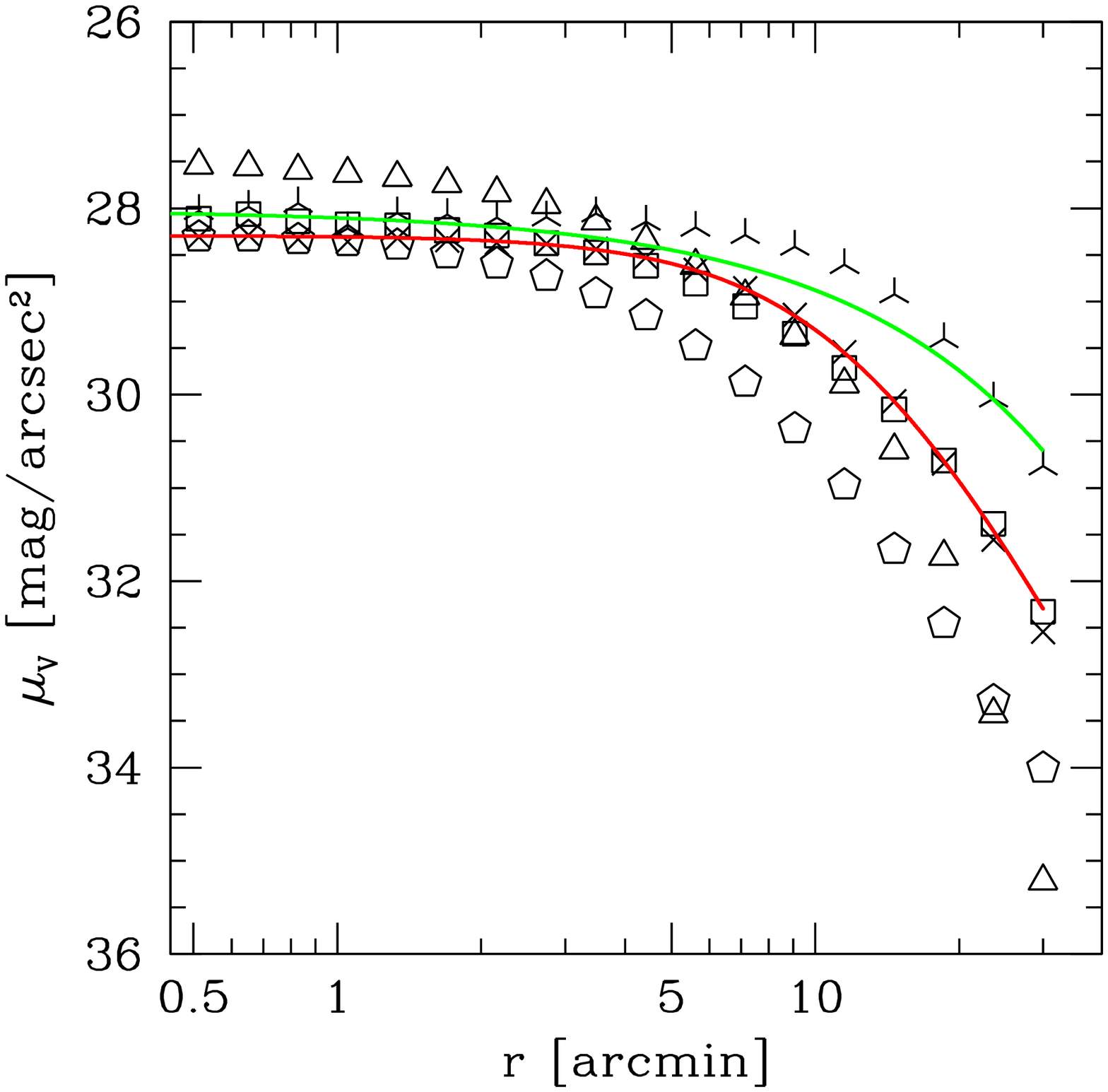}
  \epsfxsize=4.1cm
  \epsfysize=4.1cm
  \epsffile{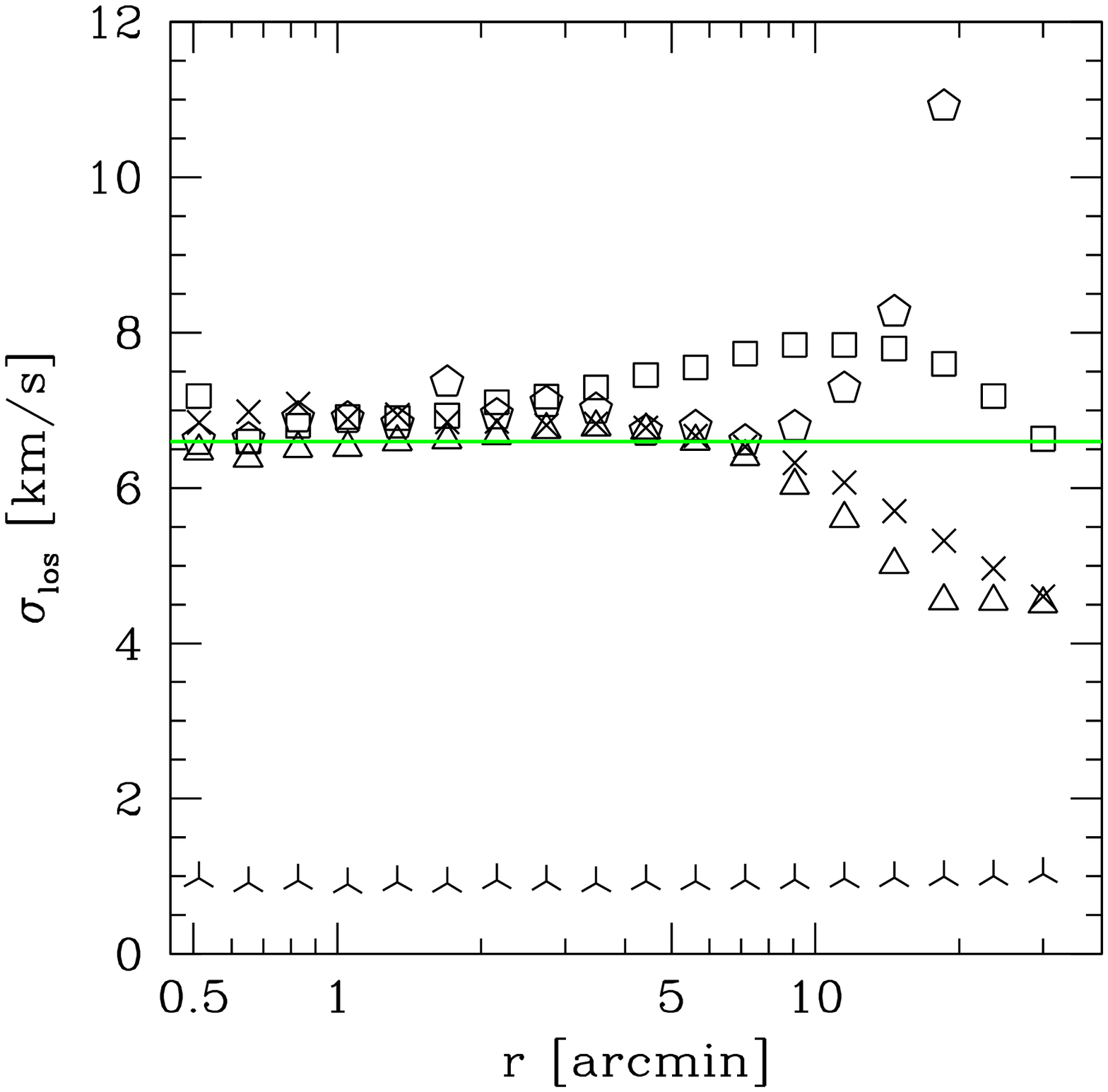}
  \caption{\textbf{Left:} Surface brightness profiles of our models.
    The upper (green) line shows the exponential fit to the observed
    data, while the lower (red) line shows the Plummer fit.
    \textbf{Right:} Velocity dispersion profiles of our models.  The
    horizontal (green) line shows the observed value of the central
    velocity dispersion.  The symbols are: Case A: tri-pods; case B:
    crosses; case C1: triangles; case C2: squares; case D: pentagons.
    Models on orbit (3) are omitted.}
  \label{fig:surfsig}
\end{figure}

\begin{figure}
  \centering
  \epsfxsize=7cm
  \epsfysize=7cm
  \epsffile{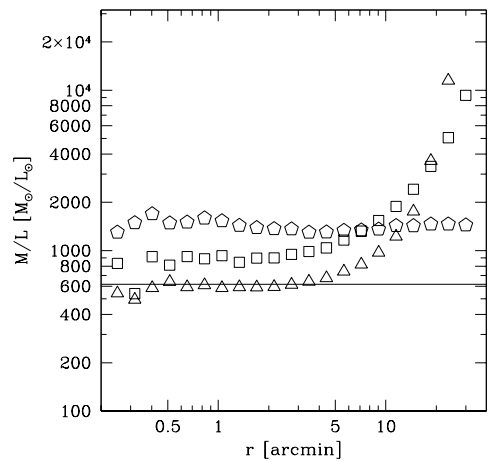}
  \caption{Radial variation of the mass-to-light ratios in our models.
    The horizontal line shows the constant value of case B (mass
    follows light).  Triangles represent case C1, squares represent
    case C2 and pentagons are for case D.  To convert the mass in
    luminous matter in our simulations to a luminosity for the stellar
    component we assume a mass-to-light ratio of two.  Models of orbit
    (3) are omitted.} 
  \label{fig:mlratio}
\end{figure}

\begin{table}
  \centering
  \caption{Parameters of the Boo models at the present epoch.  The
    columns give the total mass within a field of view of one square
    degree ($\sim 1.1$kpc), the luminous mass within the field of
    view, the mean mass-to-light (M/L) ratio, the central
    mass-to-light ratio and the central velocity dispersion.  For the
    simulation without dark matter the M/L-ratio quoted is in brackets
    and denotes the value which would be inferred by applying the
    virial theorem.} 
  \label{tab:res}
  \begin{tabular}{cccccc} \hline
    model & total mass & lumin. mass & $\overline{M/L}$ & $M/L_{0}$ &
    $\sigma$ \\  
    & [$M_{\odot}$] & [$M_{\odot}$] & & & [km\,s$^{-1}$] \\ \hline
    A & $1.1 \times 10^{5}$ & $1.1 \times 10^{5}$ & ($17$) & --- &
    $1.0$ \\ 
    B & $1.3 \times 10^{7}$ & $4.2 \times 10^{4}$ & $620$ & $620$ &
    $6.5$ \\ 
    C1 & $2.7 \times 10^{7}$ & $3.0 \times 10^{4}$ & $1800$ & $550$ &
    $6.5$ \\ 
    C2 & $6.7 \times 10^{7}$ & $3.9 \times 10^{4}$ & $3400$ & $800$ &
    $6.5$ \\ 
    D & $1.1 \times 10^{7}$ & $1.4 \times 10^{4}$ & $1400$ & $1400$ &
    $6.5$ \\ 
    A-3 & $2.3 \times 10^{5}$ & $2.3 \times 10^{5}$ & ($32$) & --- &
    $2.0$ \\ 
    B-3 & $1.3 \times 10^{7}$ & $3.3 \times 10^{4}$ & $800$ & $800$ &
    $7.0$ \\ 
    C1-3 & $1.1 \times 10^{7}$ & $2.9 \times 10^{4}$ & $740$ & $600$ &
    $5.5$ \\ 
    C2-3 & $2.7 \times 10^{7}$ & $3.6 \times 10^{4}$ & $1500$ & $1000$ &
    $6.5$ \\ 
    \hline 
  \end{tabular}
\end{table}

\subsection{The `star cluster' case (A)}
\label{sec:case-a}

In this simulation we assume that the elongation of the outer stellar
iso-density contours of Boo represents the on-set of tidal tails.  This
requires that the final model has a tidal radius of about $250$~pc
which, given the observed luminous mass, can only occur if the
enclosed mass contains no dark matter contribution.  We start with a
single Plummer sphere of mass $8 \times 10^{5}$~M$_{\odot}$ and Plummer
radius (half-light radius) of $202$~pc and truncate the distribution
at $500$~pc (see Table~\ref{tab:models}).  We choose orbit (4) for
this model because it fits the line-of-sight path traced by the tidal
tails and is not extreme in any sense.  The final object has a bound
mass of $7.6 \times 10^{4}$~M$_{\odot}$ and a mass (including unbound
material) of $1.1 \times 10^{5}$~M$_{\odot}$ lies within a field of
view of one square degree.  The object is strongly tidally distorted
and elongated, although the iso-density contours remain quite smooth
(see Fig.~\ref{fig:surf-pix}, top left panel).  The radial surface
density distribution reproduces the observed profile reasonably well
(see Fig.~\ref{fig:surfsig} left panel) but the velocity dispersion of
the final object is only of the order of $1$~km\,s$^{-1}$ (see
Fig.~\ref{fig:surfsig} right panel).  In fact, the remaining bound
mass has an even lower dispersion ($\approx 0.5$~km\,s$^{-1}$) but we
observe an enhanced value in projection due to the unbound stars
surrounding the object.  The remnant exhibits no significant velocity
gradients across the main body of the system.  As expected, there is
an offset of about $2$~km\,s$^{-1}$, or roughly twice the internal
velocity dispersion, between the mean velocities of the leading and
trailing tails.

Since this model can not reproduce the observed velocity dispersion of
Boo, even when unbound stars along the line of sight are included, we
conclude that a dark matter-free system is ruled out as a progenitor of
Boo.  Increasing the stellar mass of the progenitor would increase the
velocity dispersion but would also reduce the level of tidal
disturbance.  As a result, the remnant would contain too much stellar
mass and would not show evidence of tidal tails.

As a final possible avenue to obtain an inflated velocity dispersion
(in projection) for a remnant without dark matter, we rotate the model
(A) remnant so that the tidal tails lie along the line of sight. This
configuration would be expected to maximise the observed velocity
dispersion -- \citet{kroupa97} previously used such preferred
alignments to argue that the dynamical masses of the larger dSphs were
over-estimated. In the case of Boo, we note that its narrow horizontal
branch would argue against a significant line-of-sight extension in
this system \citep[see the colour magnitude diagram in Fig.~2
of][]{be06}. Even given this optimal alignment, however, we find that
the projected velocity dispersion of the remnant is smaller than
$2$~km\,s$^{-1}$. This adds further weight to our conclusion that all
viable progenitors of Boo must have contained dark matter.

\subsection{(Dark plus luminous) mass-follows-light: case (B)}
\label{sec:case-b}

This model is a simplified representation of a two-component
progenitor with a cored luminous distribution together with a cored
dark matter halo of the same scale-length and profile.  In this
scenario, the initial object is a Plummer sphere with a mass of $1.5
\times 10^{7}$~M$_{\odot}$.  It has a Plummer radius of $200$~pc and
the distribution is truncated at $2$~kpc (see also
Table~\ref{tab:models}).  At the present epoch, the remnant mass is
about $1.3 \times 10^{7}$~M$_{\odot}$, a value which agrees closely
with our simple mass estimate from \S~\ref{sec:inimod}.  The central
line-of-sight velocity dispersion is $6.5$~km\,s$^{-1}$ which agrees
with the observed value and the final scale-length is $13$~arcsec as
observed.  To reproduce the radial surface-brightness profile and the
total luminous mass, a mass-to-light ratio of $620$ is required (see
Fig.~\ref{fig:surf-pix} top right panel).  The resulting surface
brightness and velocity dispersion profiles are shown as crosses in
Fig.~\ref{fig:surfsig}.  The surface brightness profile is very close
to the observed Plummer fit to the profile of Boo, which is
unsurprising given that the inner regions of this model are shielded
from the effects of tides.  We note, however, that at larger radii
(not visible in the field of view shown in Fig.~\ref{fig:surf-pix})
this model exhibits the on-set of tidal tails. 

One might also speculate that the dark matter halo of Boo's progenitor
was cored but with a scale-length much larger than that of the light.
In such a model, dark matter at large radii would be tidally stripped
at earlier times.  As a result, the scale-lengths of the dark and
luminous matter distributions would become more similar over time,
leading to a model which was closer to mass-follows-light at the
present epoch.  We therefore conclude that a cored halo model is a
plausible progenitor for Boo and, further, that a total
mass-follows-light model is consistent with the currently available
data. 

As we discussed earlier, this model does not reproduce the elongation
of Boo but remains spherical, having the same luminous mass and
scale-length as the observations.  Since the inner regions of this
model (corresponding to the volume probed by the observed stellar
distribution) are almost unaffected by the external tidal field, we
expect that our results would change only marginally if we used an
elongated model with the same mass and mean scale-length at the start of
the simulation.  However, although this would not strengthen our
conclusions regarding the viability of this scenario, it would greatly
increase the computational effort required to identify suitable
progenitors as we would have to constrain two additional free
parameters (orientation angles) unrelated to the important physical
structure of Boo in order to match the observed orientation of Boo. 

Both model A and model B compare well with \citet{jo02}, probing just
much smaller scales, even though in model B features like isophotal
twist and the break radius are beyond the field of view of
Fig.~\ref{fig:surf-pix} and~\ref{fig:surfsig}.  It is the 'diffuse
light' around model A which boosts its 'measured' velocity dispersion
by a factor of two.

\subsection{Extended dark matter: case (C)}
\label{sec:case-c}

\noindent {\bf C1:} In this simulation, the dark matter halo and the 
luminous matter have the same scale-length but differing initial
radial structure and radial extent (see Table~\ref{tab:models}).  In
order to match the high central velocity dispersion, this model
requires a concentrated halo, with a correspondingly low total mass.
As a result, halo material can be easily stripped from the outer parts
of the system.  At the present epoch in our simulation we find
$3 \times 10^{4}$~M$_{\odot}$ of luminous matter in the field of view
(see Fig.~\ref{fig:surf-pix} middle left panel) and $2.7 \times
10^{7}$~M$_{\odot}$ in dark matter, which agrees approximately with
our estimate from Section~\ref{sec:inimod}.  The average M/L-ratio
within one square degree is $1800$.  If we analyse the radial
dependence of the M/L-ratio (see Fig.~\ref{fig:mlratio}) we see that
in the very centre the value is about $550$ and therefore similar to
our mass follows light case (B) and it increases only in the very
outer parts where virtually no stars are present.  At the present
epoch in the simulations the stellar distribution is slightly
elongated at larger radii (outside our field of view) although we
started with a perfectly spherical distribution.  This shows that a
lot of dark matter in the outer parts has been stripped away, and the
outer luminous part of the satellite is feeling the tidal forces of
the Milky Way.  The final radial surface brightness profile is similar
to that of Boo, although the central value is slightly higher.  As
with model (B), the elongation of the satellite at radii less than
$0.5^\circ$ is not reproduced - an initially flattened progenitor
would probably yield a remnant with a density distribution more
similar to that of Boo.  The velocity dispersion within $10^\prime$
matches the observed value, although it falls off at larger radii.\\ 

\noindent {\bf C2:} In this scenario, we have an initial model 
where the luminous matter is represented by a Hernquist sphere
embedded in a significantly more extended NFW halo.  The initial
scale-length of the halo is larger than the initial size of the
luminous distribution (see Table~\ref{tab:models}).  This model
reproduces the high velocity dispersion of Boo but extends to much
larger radii than the observed object.  By the end of simulation, we
find $3.9 \times 10^{4}$~M$_{\odot}$ of luminous matter within one
square degree but $6.7 \times 10^{7}$~M$_{\odot}$ of DM.  These values
imply a mean M/L-ratio of more than $3400$, which would be the highest
claimed ratio within the region probed by the stars in any Milky Way
satellite so far, but are comparable to those derived in NFW-based
modelling by \citet{pmn07}.   As Fig.~\ref{fig:mlratio} shows, in the
innermost regions the M/L-ratio of this model is about $800$ but it
increases exponentially with radius up to a value of $\approx 9000$.

Due to its massive, extended halo, this model has no tidal tails
inside the optical radius.  However, the surface brightness profile
closely matches the observed profile (see Fig.\ref{fig:surfsig}), and
the high central velocity dispersion is also reproduced.  As in the
previous models, we do not account for any initial ellipticity of Boo.

We conclude that if we were to allow some intrinsic elongation either
of our NFW halo scenarios could account for the observed properties of
Boo.  


\subsection{Dark matter \& tails: case (D)}
\label{sec:case-d}

\begin{figure}
  \centering
  \epsfxsize=7cm
  \epsfysize=7cm
  \epsffile{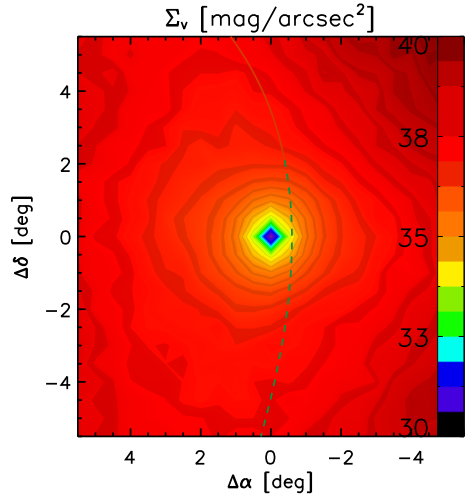}
  \caption{Surface brightness distribution over a larger area,
    complementary to Figure~2, for model (D) in which the progenitor
    is on an extreme orbit about the Milky Way.} 
  \label{fig:b123ext}
\end{figure}

In this model we attempt to retain the assumption that the observed
elongation of Boo is the onset of tidal tails in  a dark
matter-dominated model.  Therefore, we place the progenitor on a more
eccentric orbit.  For this scenario to be viable, the perigalacticon
of the orbit must be sufficiently close to the Galactic Centre that
the tidal radius of Boo shrinks to a value such that even the
deeply-embedded luminous matter is affected by the external field.  In
fact, we find that the simulation does not show tidal distortion on
small scales, although some hint of tails are noticeable on larger
scales (see Fig.\ref{fig:b123ext}).  As in our other dark
matter-dominated models, the velocity dispersion of Boo is
reproduced.  However, the evolution leads to a surface brightness
profile which falls off somewhat more steeply than the observed
profile.  In the field of view, we find $1.4 \times
10^{4}$~M$_{\odot}$ in stars and $10^{7}$~M$_{\odot}$ in DM, giving an
average M/L-ratio of about $1400$.  As Fig.\ref{fig:mlratio} shows,
the M/L-ratio is roughly constant throughout the remnant. 

This simulation shows that even though the tidal radius at
perigalacticon shrinks to values well within the luminous
distribution, and some stars become unbound, the resulting stellar
remnant does not exhibit tidal tails at the present epoch.  During the 
long period of the orbit where the tidal radius is much larger than
the extent of the luminous matter, formerly unbound stars become bound
again and, apparently, redistribute themselves symmetrically about the
remnant due to phase mixing.  Given the absence of tails, we conclude
that this model is no more successful in reproducing the observations
than our other dark matter-dominated models and merely serves to
illustrate that our conclusions relating to models (B), (C1) and (C2)
would be largely unchanged if those progenitors were placed on more
extreme orbits.  

\subsection{Orbit (3) models}
\label{sec:o3}

Because neither orbit (4) nor the extreme orbit (2) are based on
cosmological assumptions we also perform a suite of simulations using
orbit (3), which is more typical of the range of eccentricities found
in cosmological simulations.  We search for initial models for our
four orbit (4) cases, namely A to C2 and compare the different models.
The initial conditions and the final results are shown in
Table~\ref{tab:models} and Table~\ref{tab:res} respectively.
Comparing the initial values, there is a clear trend that the initial
masses have to be higher to reproduce the same kind of remnant today.
One can also notice the trend that the final models have larger
central mass-to-light ratios than their corresponding models on orbit
(4).  This trend is best visible if one compares the sequence of the
models C1(-4), C1-3 and D (or C1-2).  The initial mass increases with
decreasing perigalacticon and also the final central mass-to-light
ratio increases from 550 via 600 to an astonishing value of 1400 in
the most extreme case.  

To simplify the comparison between our models, we kept the same
initial scale-lengths from our orbit (4) models in our DM-dominated
simulations for progenitors on orbit (3).  In order to shield the
luminous component from tidal disruption, we then have to make the
orbit (3) models more massive. As a result, the initial central
M/L-ratios were higher in the orbit (3) models as well as the central
velocity dispersions.  However, the halos on orbit (3) are tidally
shaped differently, because they experience a stronger tidal field and
therefore at the end of the simulation the halo scale-lengths in our
orbit (3) models differ from the ones on orbit (4).  Due to this fact,
we still see higher central values for the M/L-ratio although the
final velocity dispersions are similar.

\begin{figure}
  \centering
  \epsfxsize=7cm \epsfysize=7cm \epsffile{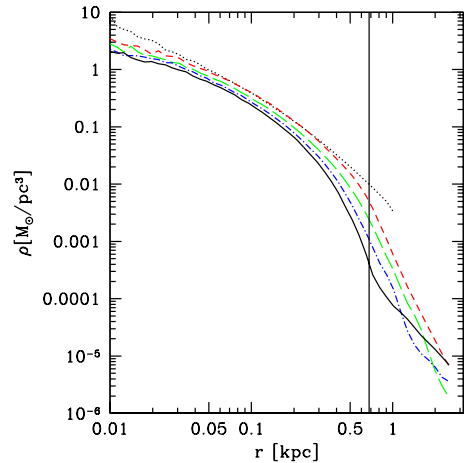}
  \caption{Evolution of the density profile of model
    C1-3.  Shown are the initial profile (dotted, black), at
    $t=2.5$~Gyr (short dashed, red), at that time the satellite is
    near its perigalacticon, at $t=5$~Gyr (long dashed, green), at
    $t=7.5$~Gyr (dashed-dotted, blue) close to its apogalacticon and
    the final profile at $t=10$~Gyr (solid, black).  Vertical line
    denotes the tidal radius at perigalacticon.}
  \label{fig:time}
\end{figure}

In our heavily dark matter-dominated simulations we find only a slight
evolution of the luminous component with time.  But the halo component
experiences a rather strong tidal shaping in its outer parts.  We show
the time evolution of the halo density profile of one of our models
(C1-3) in Fig.~\ref{fig:time}.  Still our model of the DM halo shows a
cuspy profile even after 10~Gyr of evolution, i.e.\ the influence of
the tides on the innermost parts of the satellite is rather weak.  Due
to the interplay of dark and luminous component effects as discussed
in \citet{ka04} seem not to play a strong role in our simulations.

\section{Conclusions}
\label{sec:conclus}

In this paper we have investigated a wide range of possible
progenitors and evolutionary histories for the recently discovered
Bootes dwarf spheroidal satellite galaxy of the Milky Way.  As 
we discussed in \S~\ref{sec:inimod}, although the outer iso-density
contours of the stellar distribution are elongated, an observation
sometimes interpreted as suggestive of tidal disturbance, this
interpretation is not consistent with the observed velocity dispersion
of $\approx 7$~km\,s$^{-1}$.  Our initial attempts to reproduce both
the distorted outer morphology and high velocity dispersion using
models which were close to, or slightly beyond, the point of complete 
disruption were unsuccessful.  We therefore focused on bound models
either with or without the presence of significant quantities of dark
matter. 

Our model (A) is an extended 'star cluster' model and reproduces the
surface brightness profile and distorted outer morphology of Boo
reasonably well.  As expected for a purely stellar system from our
argument in \S~\ref{sec:inimod}, however, the final velocity
dispersion is significantly smaller than that observed in Boo.  It is
therefore very unlikely that a purely stellar model could be the
origin of Boo.   

Our dark matter-dominated models models (cases (B) - (D)) can
reproduce both the surface brightness profile and the velocity
dispersion, but not the morphology of the outer stellar distribution.
This morphological limitation is because all our progenitors are
initially spherical, and their dark matter halos protect them from
significant tidal disturbance.  Even if the orbit of Boo has a very
small perigalacticon (case (D)), the remnant does not exhibit tidal
tails on scales as small as those observed in Boo.  For these models
therefore, if the elongation of the outer contours of Boo is real and
not an artifact of the sparse photometric data, then it must result
from intrinsic elongation of the progenitor system.  As stated in
\S~\ref{sec:inimod} the putative S-shape of the contours in these
cases do not represent the on-set of tidal tails but might be caused
by tidal torques acting on an intrinsically flattened system which is
at least partly rotationally supported but remains deeply embedded in
its DM halo.  Given the very low stellar mass of a faint satellite
like Boo, a dwarf disc galaxy progenitor (such as was used by
\citet{ma02} to model the Carina dSph) would either have to be orders
of magnitude lower in baryonic mass than typically observed stellar
discs, or would have to lose almost all of its initial mass (both dark
and baryonic). The latter scenario would either require an extremely
low star formation efficiency in the original disk (i.e.\ most of the
baryonic component is still in the form of gas, which gets easily
stripped by ram pressure stripping and the cosmic ultraviolet
background at high redshift as suggested by \citet{may07}), or very
strong tidal disturbance. Again we emphasise that to reproduce the
high velocity dispersion of Boo, we require a dark matter mass in the
inner regions which would preclude this level of tidally-induced mass
loss from the central parts - in our study it was not possible to
reproduce the data of Boo with a disrupted object of any kind.


On the basis of our study, we conclude that it is unlikely that the
progenitor of Boo contained no dark matter.  In fact, we found no
viable models in which dark matter was not always dominant.  External
tidal effects on the evolution of the progenitor of Boo do not
significantly affect the observed stellar velocity dispersion in the
inner regions, or the structure of the galaxy.  Our most likely models
have final DM masses in the range $1-7 \times 10^{7}$M$_\odot$.   

By changing the possible orbit of Boo we found that the closer the
orbit gets to the Galactic centre (i.e.\ smaller perigalacticon and
higher eccentricity) the larger is the mass-to-light ratio exhibited
in the undisturbed central part of the final model.

In order to understand properly the details of the specific Bootes
galaxy progenitor, however, further data are required.  First, deeper
photometric data extending to larger radii around the satellite are
essential in order to establish the full extent of the remnant and
whether there is any evidence that the elongation is tidal in origin
(e.g. whether the elongation links to larger scale tidal arms).
Secondly, a larger kinematic data set is required to study the
variation of the velocity dispersion and mean velocity with position
in Boo.  This would yield further information about the mass
distribution within the remnant.  Both these studies are currently
underway. \\

\noindent {\bf Acknowledgements}: MF, VB and DBZ acknowledge funding
by STFC.  MIW acknowledges support from a Royal Society University
Research Fellowship.  We thank Rainer Spurzem and Walter Dehnen for
useful comments.

\label{lastpage}

\end{document}